\documentclass[11pt,a4paper]{article}
\usepackage[]{times}
\usepackage{fullpage}
\usepackage{graphicx}
\usepackage{subfigure}
\usepackage{amsmath}
\usepackage{fancyhdr}
\date{}

\newcommand{\QED}{$\rule{2mm}{2mm}$}

\newcommand{\natu}{{\sf I \! N}}

\newtheorem{e-proposition}[theorem]{Proposition}

\newtheorem{e-definition}[theorem]{Definition\rm}
\newtheorem{remark}{\it Remark\/}

\title{One-parameter tetrahedral mesh generation for spheroids}

\author{
    Vitoriano Ruas$^{1,2}$
		\\[1mm]
  {\small $^{1}$ Sorbonne Universit\'es, UPMC Univ Paris 06, UMR 7190, Institut Jean Le Rond d'Alembert, France}\\
  {\small $^{2}$ CNRS, UMR 7190, Institut Jean Le Rond d'Alembert, F-75005 Paris, France.}\\[1mm]
  {\small e-mail: {\it vitoriano.ruas@upmc.fr}}}

\begin{document}
\maketitle
\thispagestyle{fancy}

\begin{abstract}

This paper deals with a simple and straightforward procedure for automatic generation of finite-element or finite-volume meshes of spheroidal domains,  
consisting of tetrahedra. Besides the equation of the boundary, the generated meshes depend only on an integer parameter, whose value is associated with the degree of refinement. More specifically the procedure applies to the case where the boundary of a curved three-dimensional domain not so irregular can be expressed in spherical coordinates, with origin placed at a suitable location in its interior. An optimal numbering of mesh elements and nodes can be accomplished very easily. Several  examples indicate that the generated meshes form a quasi-uniform  family of partitions, as the corresponding value of the integer parameter increases, 
as long as the domain is not too distorted.     
\end{abstract}

\noindent {\footnotesize \textbf{Key words:} finite elements, finite volumes, mesh generation, one-parameter, spheroid, tetrahedron}

\section{Introduction}
In the framework of the numerical solution of boundary value problems by the finite element method or the finite volume method, mesh generation plays a fundamental 
role. It has even become a crucial issue in contemporary techniques for numerical simulation such as adaptivity, for which the generation of meshes is sometimes more time-consuming than the problem solution itself. \\
In the case of three-dimensional problems widespread numerical techniques of the kind are based on partitions of equation's spatial domain into tetrahedra, by virtue of their flexibility to fit irregular shapes. Moreover the geometry of a tetrahedron conforms very well to simple algebra for both methods. This is 
the case for instance of linear finite element or vertex-centered finite volume schemes, in which the approximation of a curved domain by a polyhedron equal to the union of mesh tetrahedra does no harm in terms of accuracy. Even in the case of higher order methods this kind of geometrical approximation is acceptable, provided  
a suitable boundary condition interpolation is employed (see e.g. \cite{cmame2017}). \\
For all those reasons high quality mesh generation is a vast subject, to which an increasing number of respected specialists are steadily contributing.   
Most devoted themselves to the development of procedures for tetrahedral mesh generation as general as possible. A good survey on this topic can be found in 
\cite{Lo}. One of the pioneering work in this direction is due to Hermeline \cite{Hermeline} and to George (see e.g. \cite{George}). Several celebrated work followed included or quoted in publications such as \cite{FreyGeorge} and \cite{GHS}, to name just a few. 
In the framework of both discretization methods under consideration, the construction of Delaunay tesselations is very important. Therefore several authors contributed in this direction. In this respect we could quote for instance \cite{HermelineGeorge}, among many others. \\
It would be difficult to be exhaustive about the state-of-the-art of tetrahedral mesh generation in a single article. Our point here is that, for obvious reasons, the use of a very general procedure is not so suitable to generate tetrahedral meshes with the best properties in practical terms, when the domains has a more particular shape. If we take the example of a sphere, it is clear that a procedure especially designed for its shape would be preferable to a general one based on a triangular mesh of domain's surface, like most mesh generators use. The present contribution lies precisely in an extension of such a framework, for we deal here with a very simple procedure to generate high quality tetrahedral meshes of sphere-like domains, or equivalently spheroidal domains. Here quality means that optimal node numbering is achieved without any complex algorithm, for a mesh can be generated by inputting only a single integer parameter defining its degree of 
refinement. It also means that the mesh tetrahedra have approximately the same shape and volume, as long as the domain is not too distorted as compared to a sphere.  
Besides the usual stability and consistency requirements, the  latter property, known as quasi-uniformity or uniform regularity (see e.g.\cite{Ciarlet}), 
is sufficient to guarantee accuracy improvement of the discretization method as the mesh is refined. But more than this, quasi-uniformity is the condition under which 
important tools of the mathematical analysis of a numerical method for partial differential equations apply. This is for instance the case of inverse inequalities 
for Sobolev norms (cf. \cite{Ciarlet}). Moreover it is always handy to use simple procedures, that nevertheless attempt to distribute mesh elements in such a way that smaller elements are naturally assigned to domain's narrowest zones. This is the case of the one proposed in this paper, even though only one parameter 
determines the construction of the partition.\\   
As one can infer from the above introduction, the main limitation of our mesh generation method is the fact that it requires that the domain be not very irregular. More specifically we confine ourselves to the case where its boundary can be expressed in spherical coordinates for a suitable origin located in its interior. 
Actually in order to guarantee that the already mentioned regularity properties will hold, it is advisable to further require that the domain is star shaped with respect to all points of a sub-domain having a non negligible measure with respect to its own measure. Then any point in the interior of this sub-domain can be taken as the origin of spherical coordinates. For most practical geometries our method is designed for, the best choice of the origin is obvious, such as in the case of a sphere or an ellipsoid.\\
As we should point out, the method to be described hereafter is a non trivial three-dimensional counterpart of a one-parameter triangular mesh generation procedure studied in \cite{CMA} for star shaped two-dimensional domains. Likewise the boundary of the domains it applies to can be expressed in polar coordinates with origin at a suitable point in its interior. Very nice meshes of disks, ellipses, among less classical domains encountered in practical applications have been generated with such a procedure in several author's work (see e.g. \cite{book}).\\
Likewise its two-dimensional analog \cite{CMA}, the mesh generation method considered in this article is particularly suitable to check the order of a new discretization method, in case the equation to solve is posed in a curved domain. This is because mesh successive refinement is very easy to carry out, and a roughly uniform sequence of meshes is thus generated, as seen in the examples given in the sequel.\\  

An outline of the paper is as follows. In Section 2 we describe our one-parameter tetrahedrization procedure, by defining its vertices together with the way thay are linked together in order to form the final partition. In Section 3 we complete this description by specifying the steps allowing for the practical calculation of the vertex coordinates. In Section 4 we construct meshes of some star shaped domains in order to exemplify our tetrahedrization procedure. In particular we observe 
numerically mesh quality in terms of both refinement and domain distortion. Finally we conclude in Section 5 with a few remarks.                    

\section{Partition Description}

To begin with we consider a modification of the usual partition of a unit cube $C$ into tetrahedra, based on its subdivision into macro-tetrahedra. 
Taking the origin $O$ of a system of cartesian coordinates $x_1$, $x_2$, $x_3$ to be the center of $C$, the corresponding axes are chosen parallel 
to its edges. Those axes subdivide $C$ into eight equal cubes, each one of them corresponding to an octant of the three-dimensional space. \\
\indent Next we denote these eight cubes and octants by $C_{\mu}$ and ${\bf O}_{\mu}$, respectively, where $\mu$ is triple subscript 
$(\mu_1,\mu_2,\mu_3)$ such that $\mu_i = [sign(x_i)+1]/2$ for $i=1,2,3$, $x_i$ being any non zero value of the $i$-th coordinate in ${\bf O}_{\mu}$. \\
\indent Now referring to Figure 1 we take as a model octant ${\bf O}_{\nu}$ with $\nu=(1,1,1)$ for the purpose of this description. 
In doing so let $d=OD$ be the diagonal of $C_{\nu}$ which is a half diagonal of $C$, $d_1$, $d_2$, $d_3$ be the diagonals of the faces 
of $C_{\nu}$ intercepting at $O$, and $d_4$, $d_5$, $d_6$ be the diagonals of the faces of $C_{\nu}$ intercepting at the end $D$ of $d$. $d_1$, $d_2$, $d_3$, $d_4$, $d_5$, $d_6$ subdivide $C_{\nu}$ into six equal macro-tetrahedra, which we denote by $T_{\nu \alpha}$, where $\alpha=(\alpha_1,\alpha_2,\alpha_3)$  is another triple subscript corresponding to a permutation of ${1,2,3}$. More precisely the position of $T_{\nu \alpha}$ illustrated in Figure 1 is 
such that in each point of this macro-tetrahedron we have $x_{\alpha_1} \geq x_{\alpha_2} \geq x_{\alpha_3}$. Now given an integer parameter $p$, $p \geq 1$, we 
subdivide $C_{\nu}$ into $p^3$ equal cubes. Next we bring together the vertices of those cubes located in the interior and faces of each macro-tetrahedron 
$T_{\nu \alpha}$ by segments parallel to its six edges. In this manner a partition of $C_{\nu}$ into $6 p^3$ equal tetrahedra is generated. \\
\begin{figure}[h]
\label{fig1}
\begin{center}
\includegraphics[scale=0.45]{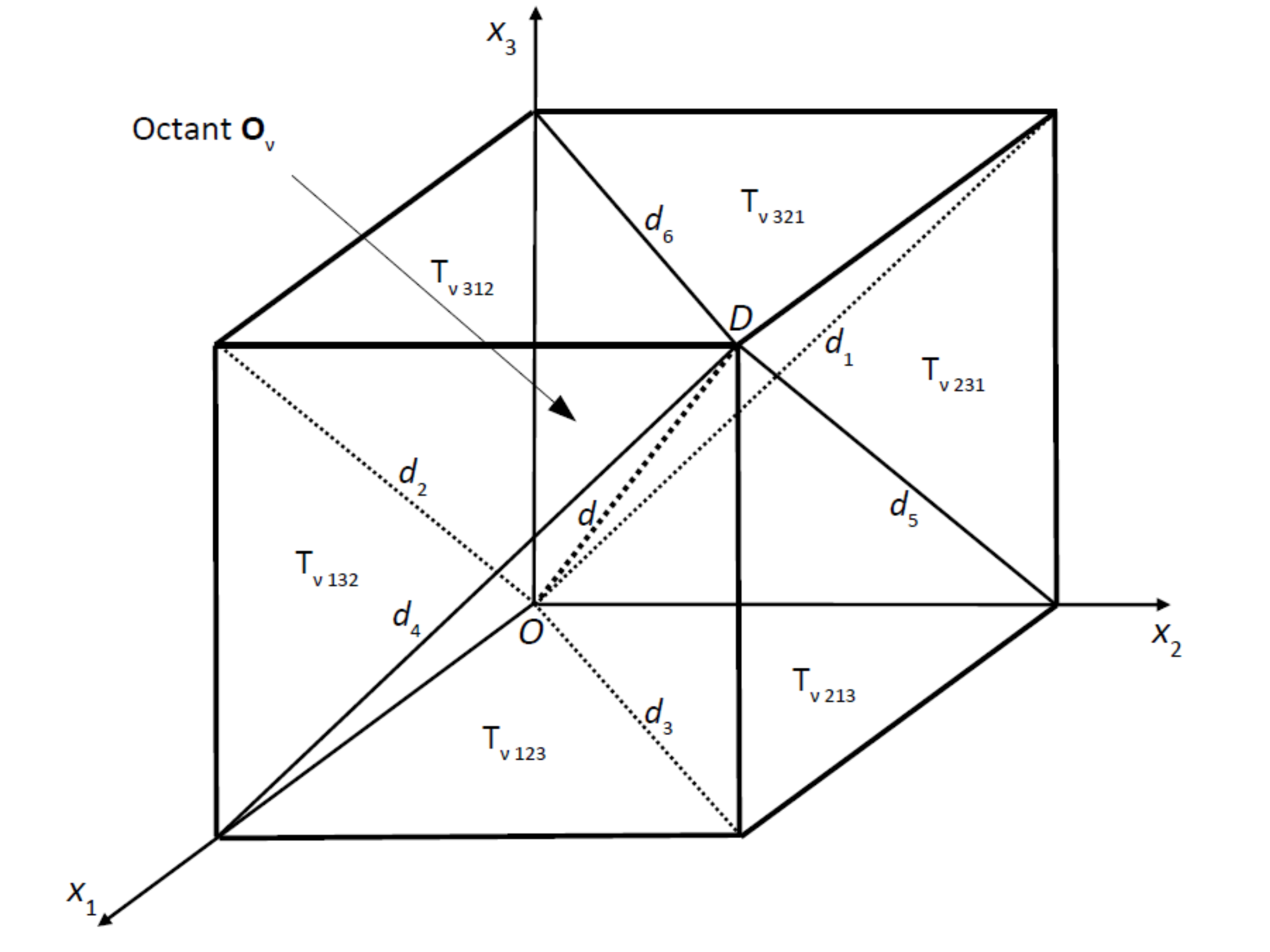}
\end{center}
\par
\caption{Cube $C_{\nu}$ and the six tetrahedra $T_{\nu \alpha}$ around its diagonal $d=OD$ it is subdivided into}
\end{figure} 
\indent Finally, taking the half diagonals of $C$ as a starting point, we proceed in the same manner for the other seven octants using symmetry, thereby generating a partition of the unit cube into $48 p^3$ equal tetrahedra. Notice that the cartesian coordinates of vertices of all tetrahedra of such a partition are of the form $(i_1/[2p],i_2/[2p],i_3/[2p])$ where the $i_k$s are integers in the interval $[-p,p]$. Furthermore it is possible to number the vertices of the partition in a structured manner from one through $(2p+1)^3$.  More precisely, for example, we can number the vertices located 
on the face given by $x_1 = i_1/(2p)$ one after the other from $i_1=-p$ up to $i_1=p$, in the usual way for squares, as shown in Figure 2 for the $(i_1+p+1)$-th face.   \\
\begin{figure}[h]
\label{fig2}
\begin{center}
\includegraphics[scale=0.45]{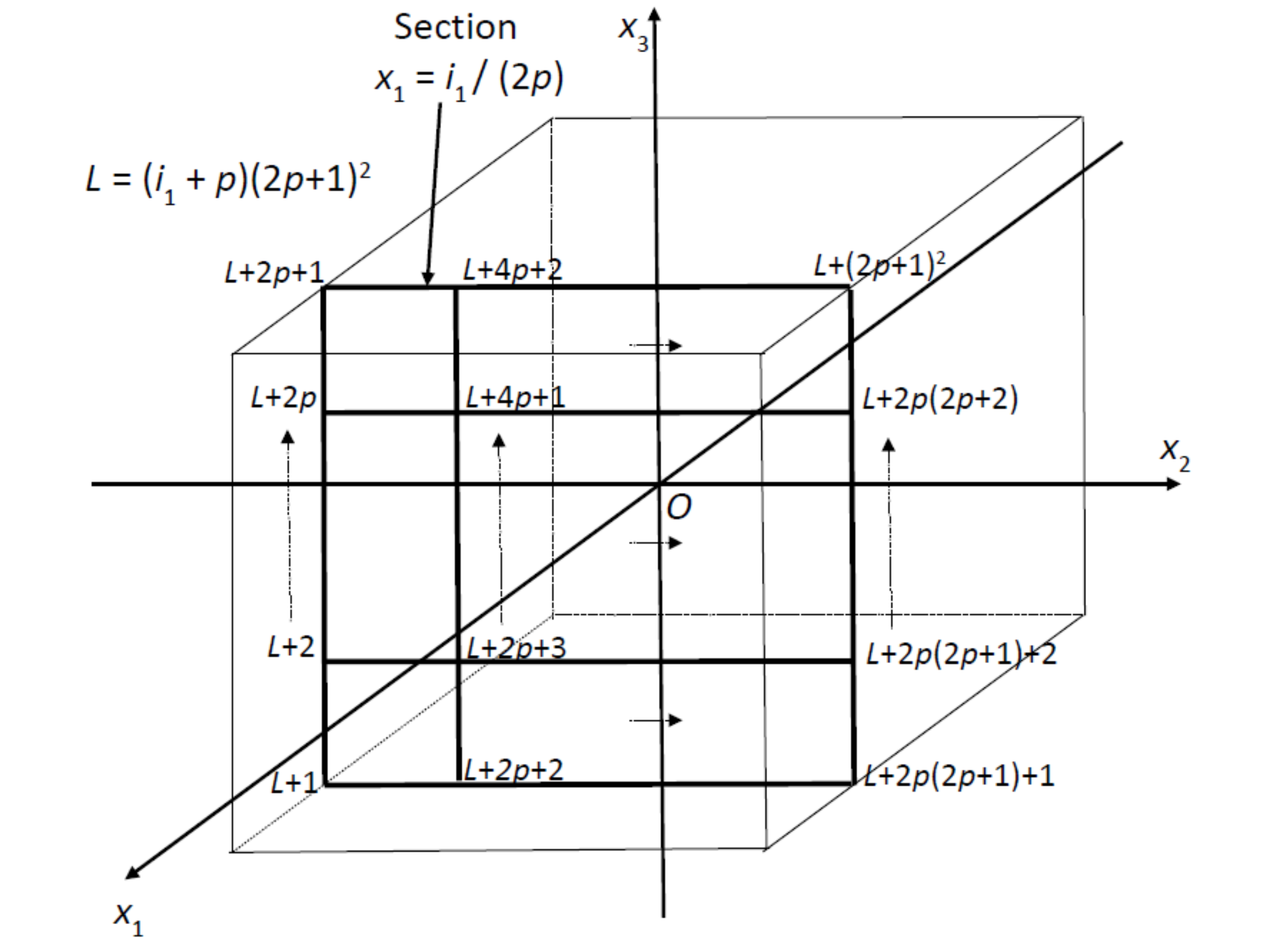}
\end{center}
\par
\caption{Numbering of the vertices on a typical section of the unit cube $C$ given by $x_1=$constant}
\end{figure} 
For the later convenience we point out that the coordinates of the vertices belonging to macro-tetrahedron $T_{\mu\alpha}$ can be written as follows:
\begin{equation}
\label{coor}
\left\{
\begin{array}{ll}
x_{\alpha_1} = (-1)^{\mu_1} \; \displaystyle \frac{-l}{2p}, & l=0,1,\ldots,p \\
x_{\alpha_2} = (-1)^{\mu_2} \; \displaystyle \frac{-m}{2p}, & m=0,1,\ldots,l \\
x_{\alpha_3} = (-1)^{\mu_3} \; \displaystyle \frac{-n}{2p}, & n=0,1,\ldots,m.
\end{array}
\right.
\end{equation}

Now let $\Omega$ be a star shaped domain of $\Re^3$ with boundary $\partial \Omega$ assumed to be of the $C^1$-class. Such an assumption is not mandatory, and is aimed at simplifying the presentation. $\partial \Omega$ is defined by an equation of the form $\rho=f(\theta,\phi)$ in spherical coordinates with origin $O$ conveniently chosen in the interior of $\Omega$. $\rho$ is the radial coordinate, $\theta$ is the azimuthal angle (or longitude) and $\phi$ (or $\varphi$, as some authors prefer) is the polar angle (or colatitude). We shall generate a partition of this domain into tetrahedra by a method 
entirely analogous to the one we just described for the unit cube $C$. The idea is to transform cartesian coordinates into spherical coordinates 
in a specific way for each one of the $48$ trihedra $\Re^3$can be subdivided into, corresponding to the tetrahedra $T_{\mu\alpha}$.   \\ 
\indent To begin with, here again we first subdivide $\Omega$ into eight octants defined by the cartesian axes, the latter being also associated with the spherical 
coordinates with the same origin $O$. Akin to the case of the cube we denote by $\Omega_{\mu}$ the subset of $\Omega$ contained in the octant ${\bf O}_{\mu}$, and take as a model a partition of $\Omega_{\nu}$ with $\nu=(1,1,1)$ defined as follows:\\
First of all we observe that $\Omega_{\nu}$ is characterized by $0 \leq \theta \leq \pi/2$ and $0 \leq \phi \leq \pi/2$, and set $\bar{\theta}= \pi/4$ and 
$\bar{\phi} = acos (\sqrt{3}/3)$. Next, referring to Figure 3, we subdivide $\Omega_{\nu}$ into six disjoint subsets $\tau_{\nu \alpha}$ quite abusively called macro-tetrahedra with three plane faces and one curved face contained in $\partial \Omega$, where the triple subscript $\alpha$ is defined as above. Notice that each one of these macro-tetrahedra correspond to the intersection with $\Omega_{\nu}$ of one of the six trihedra with vertex $O$, having an edge aligned with the line given by $\theta = \bar{\theta}$ and $\phi=\bar{\phi}$ (i.e. the with the segment $\delta=O\Delta$ in Figure 3), a second edge being a positive (cartesian) coordinate semi-axis. The third edge of anyone of such trihedra is the bisector of one of the  two quadrants formed by the above positive semi-axis and another positive coordinate semi-axis (in Figure 3 the curved thetrahdra $\tau_{\nu \alpha}$ are separated by the curved dashed lines and the segment $\delta$). \\
\begin{figure}[h]
\label{fig3}
\begin{center}
\includegraphics[scale=0.45]{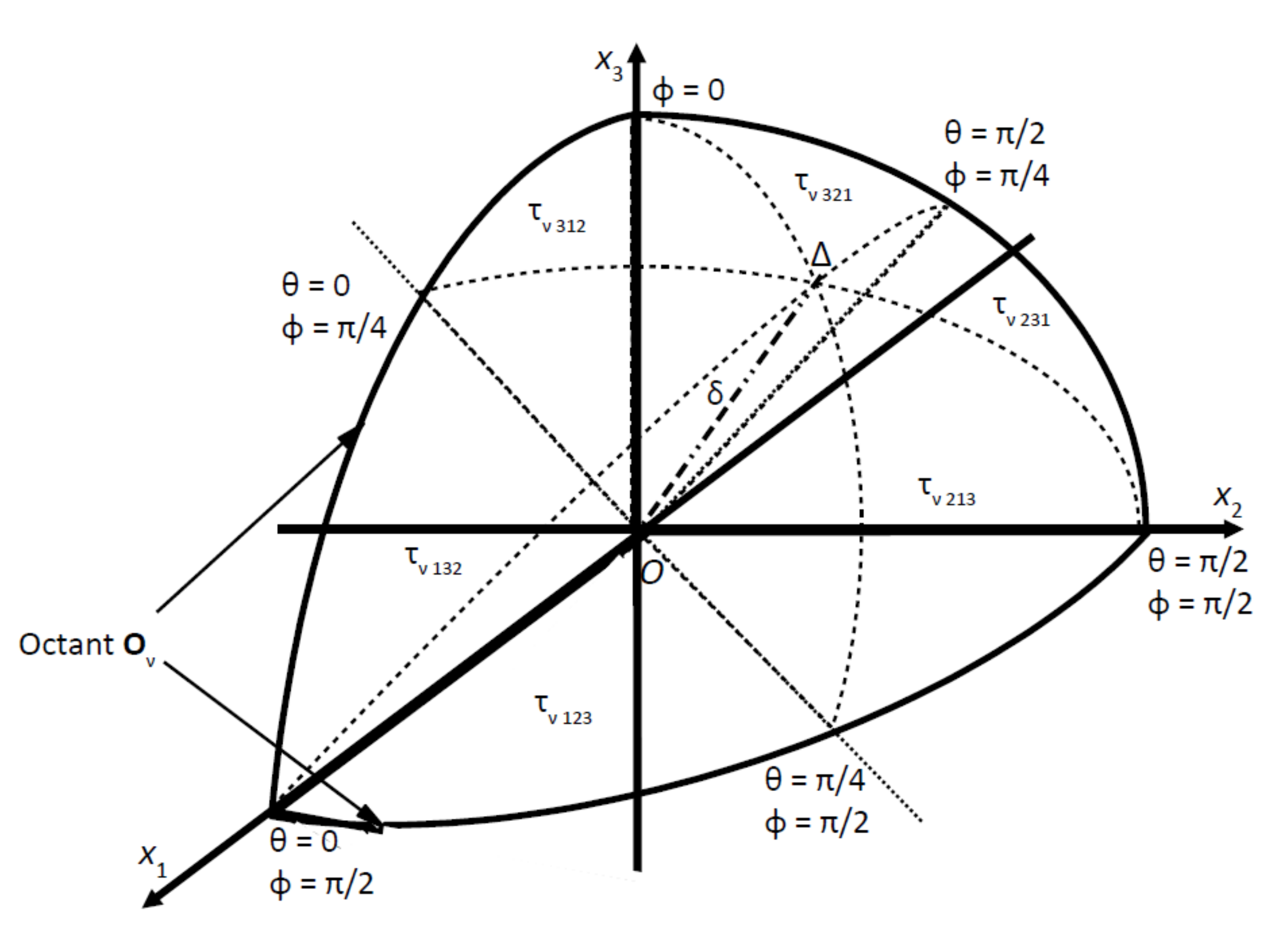}
\end{center}
\par
\caption{The curved thetrahedra $\tau_{\nu \alpha}$ in the octant ${\bf O}_{\nu}$}
\end{figure}
\indent Now let $P_{\nu \alpha1}$, $P_{\nu \alpha2}$ and $P_{\nu \alpha3}$ be the three vertices of $\tau_{\nu\alpha}$ located on 
$\partial \Omega$. Let also $(\theta_{\nu \alpha i},\phi_{\nu \alpha i})$ be the angular spherical coordinates of $P_{\nu \alpha i}$ for $i=1,2,3$. As a reference we 
take $\theta_{\nu \alpha 3}=\bar{\theta}$ and $\phi_{\nu \alpha 3}=\bar{\phi}$ for all $\alpha$ and choose $P_{\nu \alpha 1}$ to be the vertex located on 
axis $Ox_{\alpha_1}$. \\
\indent Next we consider homothetic transformations $\Omega_l$ of $\Omega$ with origin $O$ and ratio $l/p$ and let $\partial \Omega_l$ be its boundary, for $l=0,1,\ldots,p$. For each $\tau_{\nu \alpha}$ the vertices of the partition are the points $P_{\nu \alpha}^{lmn}$, for integers $m$ and $n$ 
with $0 \leq m \leq l$ and $0 \leq n \leq m$ defined in the following manner :\\
First of all we set $P^{l00}_{\nu \alpha} = l P_{\nu \alpha 2}/p$ for every $l$. Then for a given $l$ and $m \geq 1$ we denote by 
$\widehat{MON}$ the angle with vertex at the origin whose edges contain points $M$ and $N$ different from $O$, respectively. For mere convenience we call $M$ the 
"left end" and $N$ the "right end" of the angle $\widehat{MON}$ and denote by $Q^l$ the point given by $l Q/p$ for every $Q \in \partial \Omega$. 
For instance, an illustration of the points $P^l_{\nu \alpha i} \in \partial \Omega_l$, $i=1,2,3$, for $l=4$ and $\tau_{\nu 2 3 1}$ is supplied in Figure 4.   
Let also $M_{\nu \alpha}^{lm}$ and $N_{\nu \alpha}^{lm}$ be the intersection with $\partial \Omega_l$ of the polar radii that subdivide the angles 
$\widehat{P^l_{\nu \alpha 1} O P^l_{\nu \alpha 2}}$ and $\widehat{P^l_{\nu \alpha 1} O P^l_{\nu \alpha 3}}$ into $l$ equal angles in the same plane, respectively, 
for $0 \leq m \leq l$. These points are numbered from $m=0$ through $m=l$ 
from angle's left end to angle's right end. The points $P_{\nu \alpha}^{lmn}$ are the intersections with $\Omega_l$ of the polar radii that subdivide 
$\widehat{M_{\nu \alpha}^{lm} O N_{\nu \alpha}^{lm}}$ into $m$ equal angles (in the same plane) numbered from $n=0$ through $n=m$, from angle's left end to 
angle's right end. An illustration of the points $P^{lmn}_{\nu \alpha}$ is given in Figure 4 for $l=4$ and $\alpha=(2,3,1)$.\\
\begin{figure}[h]
\label{fig4}
\begin{center}
\includegraphics[scale=0.55]{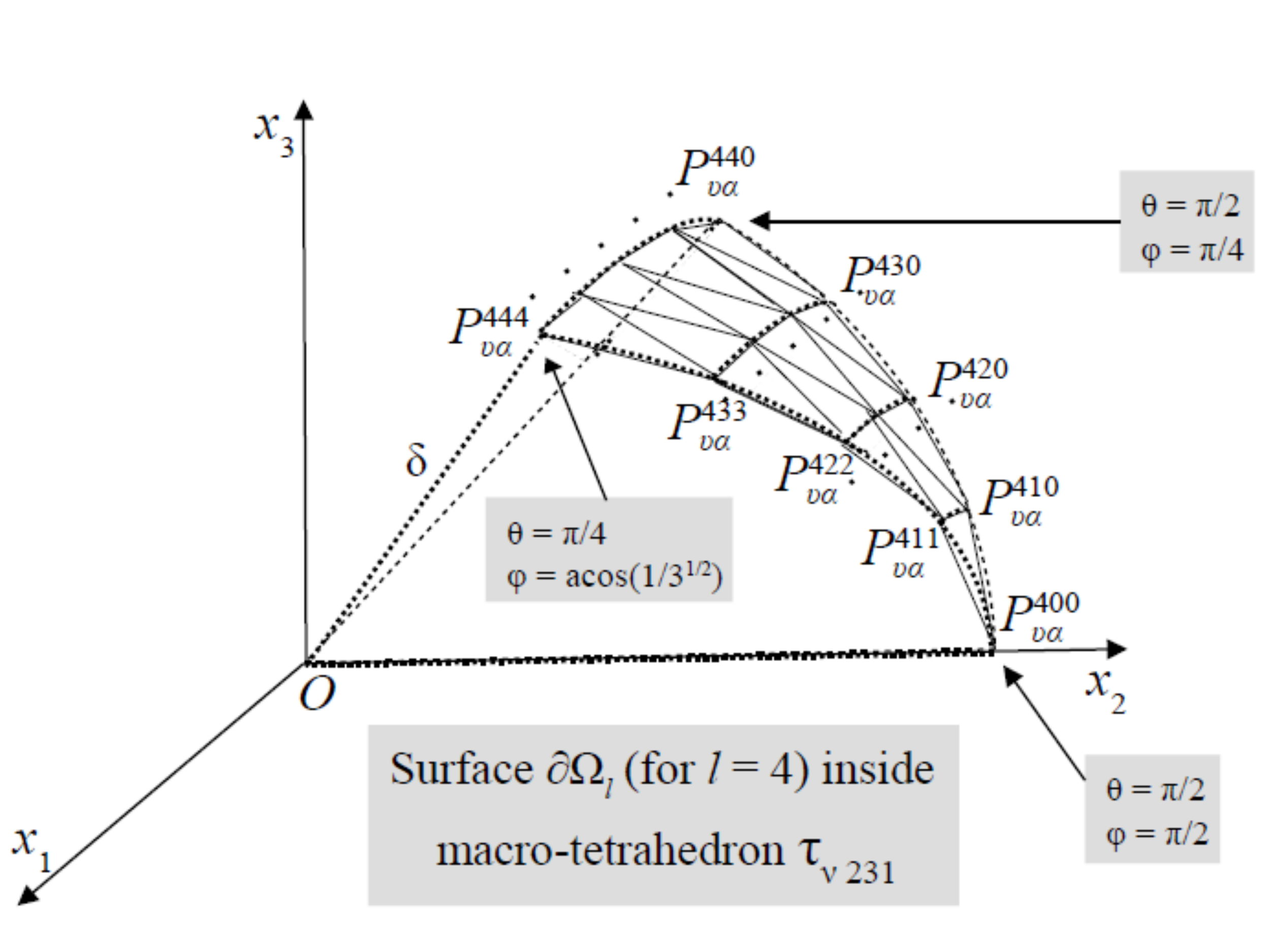}
\end{center}
\par
\caption{Points $P^{4mn}_{\nu \alpha} \in \partial \Omega_4 \cap \tau_{\nu \alpha}$ for $\alpha=(2,3,1)$ with 
$P^{400}_{\nu \alpha}=P^4_{\nu \alpha 1}$, 
$P^{440}_{\nu \alpha}=P^4_{\nu \alpha 2}$, 
$P^{444}_{\nu \alpha}=P^4_{\nu \alpha 3}$}
\end{figure}
Finally we construct a partition of the entire domain $\Omega$ by application of the principle we have just described in a symmetrically analogous manner to the other seven spatial octants. This means that for each octant ${\bf O}_{\mu}$ we define six (curved) macro-tetrahedra $\tau_{\mu \alpha}$ in such a way that the axis 
$Ox_{\alpha_1}$ contains a straight edge of $\tau_{\mu \alpha}$ for each $\mu$, and a face of the same $\tau_{\mu \alpha}$ is contained in the plane $x_{\alpha_3}=0$. Similarly, $\forall \mu$ and $\forall \alpha$, $P_{\mu \alpha 3}$ is the point of $\partial \Omega$ whose angular spherical coordinates 
$(\theta,\phi)$ are given respectively by:
\[ \theta_{\mu \alpha 3} = 5 \pi/4 + (\mu_1-\mu_2-2\mu_1 \mu_2)\pi/2 \mbox{ and } \phi_{\mu \alpha 3} = (2 \mu_3-1) \bar{\phi} \; \forall \alpha, \]
while $P_{\mu \alpha 1}$ is the point of $\partial \Omega$ located on the axis $Ox_{\alpha_1}$. Then the vertices of the tetrahedra of the final partition are determined in the same manner as for the macro-tetrahedron $\tau_{\nu \alpha}$. 
In Figure 4 the vertices of the partition belonging to $\tau_{\nu \alpha}$ with $\alpha=(2,3,1)$ located on $\partial \Omega_l$, are illustrated for $l=4$. \\           
\indent Once the vertices of the partition are known there are different possibilities to define the final tetrahedrization of $\Omega$ within each 
curved macro-tetrahedron $\tau_{\mu \alpha}$. We will chose the following one that ensures mesh compatibility on the interfaces of the $\tau_{\mu \alpha}$s, as seen 
hereafter. First of all we refer to the already described tetrahedrization of the unit cube $C$. Recalling the expressions (\ref{coor}) of the vertex coordinates 
for that partition, we can immediately establish a one-to-one correspondence between them and the above defined vertices of the intended 
tetrahedrization of $\Omega$. More specifically this means that $P_{\mu \alpha}^{lmn}$ corresponds to the point of the unit cube whose cartesian coordinates are given by (\ref{coor}). It follows that, if we assign to the  $P_{\mu \alpha}^{lmn}$s the same number as its counterpart in $C$ we can generate the tetrahedra in the partition of $\Omega$, by simply defining their edges as the segments whose ends carry the same pair of vertex numbers as for the edges of the tetrahedra in 
the partition of the unit cube. \\
\indent It is clear that in the above manner we construct a tetrahedrization of $\Omega$ consisting of $48 p^3$ elements. These tetrahedra can obviously be numbered 
in the same way as for the unit cube $C$, i.e., the number of each tetrahedron in the partition of $\Omega$ is the same as the number of an element in the partition of $C$, whenever the numbers of their four vertices coincide. \\
\indent To conclude we observe that the faces of the tetrahedra contained in the plane interfaces of two contiguous macro-tetrahedra form a triangulation 
of a plane sector with angle equal to $\pi/4$. It turns out that such triangulation coincides with the one constructed by the procedure for two-dimensional star shaped domains proposed in \cite{CMA}, as illustrated in Figure 5 for $p=5$. \\
\begin{figure}[h]
\label{fig5}
\begin{center}
\includegraphics[scale=0.55]{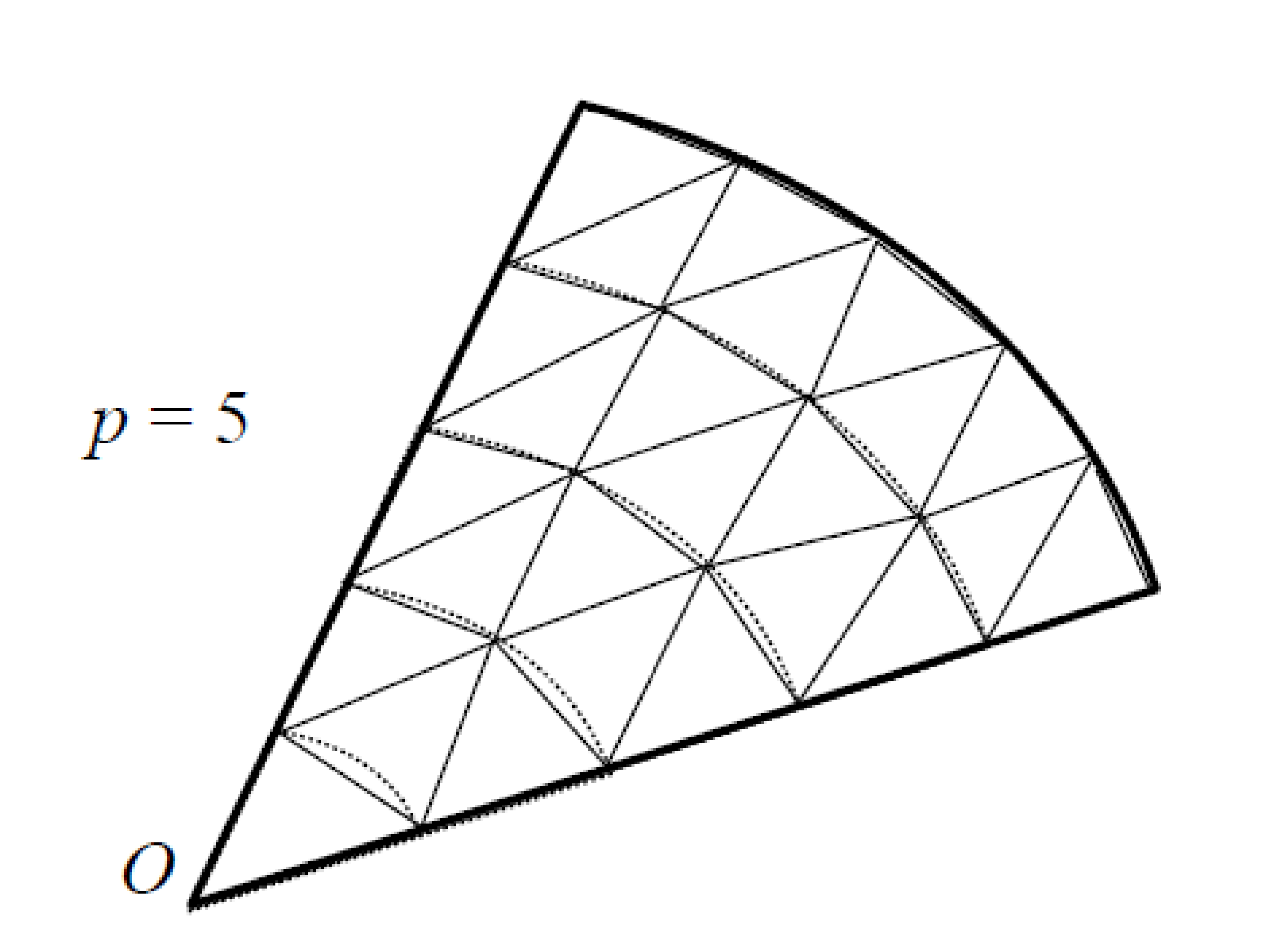}
\end{center}
\par
\caption{Traces of the mesh tetrahedra on a face common to two contiguous macro-tetrahedra for $p=5$}
\end{figure}     

\section{Determining vertex coordinates}

It is possible to set up a method for calculating the cartesian coordinates of every vertex of the partition into tetrahedra of a spheroidal domain $\Omega$ described in the previous section, given its number $k$, with $1 \leq k \leq (2p+1)^3$. This is a simple by-product of the numbering of the octants ${\bf O}_{\mu}$ and macro-tetrahedra $\tau_{\mu \alpha}$ advocated therein, together with the integer superscripts $l$, $m$, $n$, which can be associated with the three spherical coordinates as seen below. \\
First of all we determine the three integers $k_1$, $k_2$, $k_3$ with $1 \leq k_i \leq 2p+1$ for $i=1,2,3$, that fulfill  
$k = k_3 + k_2 (2p+1) + k_1 (2p+1)^2$. In doing so the values of $\mu_1$, $\mu_2$ and $\mu_3$ are given by 
\[ \mu_i = \displaystyle N \left( \frac{k_i}{p+1} \right), \]
where $N(x) : = \sup\{n \; | \; n \in \natu, n \leq x \}$. Next setting $i_j = | k_j - p - 1 |$ for $j=1,2,3$, $\alpha$ is determined by ordering the $i_j$s in such a way that $i_{\alpha_1} \geq i_{\alpha_2} \geq i_{\alpha_3}$.\\
Now all that is left to do is to subdivide the angles in the way described in Section 2, to obtain its cartesian coordinates according to the following recipe: 
Let $M$ and $N$ be two points whose angular spherical coordinates are $(\theta_M,\phi_M)$ and $(\theta_N,\phi_N)$, respectively, $\beta$ be the measure of 
$\widehat{MON}$ and $\beta_M=r \beta/q$, $\beta_N=(q-r)\beta/q$ for two integers $r$ and $q$ satisfying $q \geq r \geq 0$ and $q>0$. In practice we will have 
either $q=l$ and $r=m$ or $q=m$ and $r=n$, for $1 \leq l \leq p$ and $1 \leq m \leq l$. Now the components $u$,$v$,$w$ of the unit vector $\overrightarrow{OU}$  oriented like the polar radius in the plane of $\widehat{MON}$ that subdivides this angle into two angles (in the same plane) contiguous to $M$ and $N$, with the complementary measures $\beta_M$ and $\beta_N$ respectively, satisfy the following equations:
\begin{equation}
\label{system}
\left\{
\begin{array}{l}
a_M u + b_M v + c_M w = d_M \\
a_N u + b_N v + c_N w = d_N  \\
u^2 + v^2 + w^2 = 1, \\
\mbox{where} \\ 
d_M = cos \beta_M, \; c_M = sin \phi_M, \; b_M = cos \phi_M sin \theta_M, \; a_M = cos \phi_M cos \theta_M, \\ 
d_N = cos \beta_N, \; c_N = sin \phi_N, \; b_N = cos \phi_N sin \theta_N, \; a_N = cos \phi_N cos \theta_N.
\end{array}
\right. 
\end{equation}        
The two first equations express the fact that the point $U=(u,v,w)$ is located on the surfaces of two cones with vertex at the origin, axes $OM$ and $ON$, and apertures equal to $2 \beta_M$ and $2 \beta_N$, respectively. Noticing that there is only one point located simultaneously on the surface of the unit ball centered at the origin and on the surfaces of both cones, system (\ref{system}) has a unique easy-to-compute solution. Finally, once the components $(u,v,w)$ of the unit vector in the direction $\overrightarrow{OP}$ are determined, where $P$ is a generic notation for the vertex $P^{lmn}_{\mu \alpha}$ whose coordinates we are calculating, we can compute associated spherical coordinates $(\theta,\phi)$. Then from the radial coordinate of $P$ given by $\rho = l f(\theta,\phi)/p$, we can immediately determine its cartesian coordinates. 
\begin{remark}
In practice it is not necessary to solve (\ref{system}) as a non linear system of algebraic equations. This is because it necessarily has a unique solution. 
Therefore, after elimination of two unknowns using the first two equations of (\ref{system}), we come up with a quadratic equation  
$at^2+ b t + c = 0$ for the remaining unknown $t$, which may be either $u$, $v$ or $w$. Disregarding round-off errors the (unique) solution of this equation must be $t=-b/(2a)$ and we are done. \QED    
\end{remark}
    
\section{Quality assessment}

In this section we assess the quality of the meshes generated by the procedure described in the previous sections for some representative spheroidal domains. More precisely we use two types of data to work this out. The first one allows to check, on the basis of two different metrics, the quality of meshes with a fixed $p$, of domains with decreasing aspect ratios, in such a way that the mesh parameter $h$ also remains fixed. 
A second type of data refers to a given domain, for whose meshes we observe the evolution of the same metrics as above, as $p$ increases at the same rate as $h^{-1}$. Denoting by $V(T)$ the volume of a mesh tetrahedron $T$, among the most used metrics (see e.g. \cite{ZHB}) we choose both the ratio $r_{vr}$ 
given by
\begin{equation}
\label{VR} 
r_{vr} = \displaystyle \left[ \displaystyle \min_{T} V(T) \left/ \displaystyle \max_{T} V(T) \right]^{1/3} \right. ,
\end{equation} 
and the minimum $r_{jl}$ of the normalized Joe-Liu parameter $p_{jl}(T)$ over all mesh elements $T$. Referring to \cite{ZHB}, and denoting by 
$e_i(T)$ the six edges of $T$ for $i=1,2,3,4,5,6$, our normalization consists of taking the square root of the usual value of this parameter 
(cf. \cite{ZHB}), that is,
\begin{equation}
\label{JL} 
r_{jl} = \displaystyle \min_{T} p_{jl}(T) \; \mbox{ with } \; p_{jl}(T) = \displaystyle 2 \times 3^{5/6} [V(T)]^{1/3} \left/
\displaystyle \left[\sum_{i=1}^6 | e_i(T)|^2 \right]^{1/2} \right. .
\end{equation}

Before starting the evaluation of our mesh generation procedure it is important to list some facts about the above metrics. \\
First of all $r_{vr}=1$ corresponds to a uniform mesh, such as the one of the unit cube described in Section 2. It should also be noted that $p_{jl}(T) \leq 1$ and $p_{jl}(T) = 1$ in case $T$ is equilateral. 
The volume ratio criterion will enable us to exhibit (or not) the quasi-uniformity property of thus generated families of meshes. On the other hand the  
metrics based on the Joe-Liu parameter can indicate that a family of meshes is (shape) regular in the sense of \cite{Ciarlet}, but by no means whether or not it is 
quasi-uniform.   
    
\subsection{Mesh quality for ellipsoids with decreasing aspect ratios}
Here we consider $\Omega$ to be the ellipsoid given by $x_1^2/a^2+x_2^2/a^2+x_3^2 \leq 1$. We will take $a \leq 1$, and more particularly we will let 
the value of this parameter decrease in such a way that we will gradually switch from a sphere for $a=1$, to a cigar-shaped domain with $a=0.1$. Owing to symmetry the mesh will be generated only in the octant ${\bf O}_{\nu}$.\\
The least to be expected of the procedure being checked is that it engenders nicely regular meshes of a sphere. Thus to begin with we take $a=1$, and display 
in Table 1 the evolution of the metrics $r_{vr}$ and $r_{jl}$ as $p$ increases. The number of tetrahedra in each mesh equal to $6p^3$ is also supplied. The figures clearly indicate that the meshes behave roughly like uniform meshes of a unit cube, for which both metrics are invariant with $p$. Moreover the mesh elements are not so different from each other, since both their volumes and their shapes are rather close, as indicated by $r_{vr}$ and $r_{jl}$ respectively.\\    
As for ellipsoids, we display in Table 2 the evolution of metrics $r_{vr}$ and $r_{jl}$ for $a$ equal to $1.0$, $0.8$, $0.6$, $0.4$, $0.2$ and $0.1$ for two different meshes, namely, for $p=10$ and $p=50$. Good news here is the low sensitivity to mesh refinement of both metrics. As for the shapes and volumes a steady but relatively moderate deterioration of rates is observed, as the aspect ratio $a$ decreases. However this is more than natural, taking into account the 
significant variation of domain's shape.     

\subsection{Mesh quality for domains with pronounced  concavities}
In these experiments $\Omega$ is the domain given by 
\begin{equation}
\label{concav}
\Omega= \{(\rho,\theta,\phi) \; | \rho \leq [1+ b cos(4 \theta)][1 + b cos(4 \phi)]\},
\end{equation} 
for a parameter $b \in [0,0.4]$. Notice that if $b=0.4$ the value of the polar radius $\rho$ ranges between $0.36$ and $1.96$ within rather small sub-domains of $\Omega$. Whatever the case, for $b>0$, $\Omega$ has boundary concavities that become sharper as $b$ increases. Akin to  
Subsection 4.1 and for the same reason, only the octant ${\bf O}_{\nu}$ will be taken into account in the mesh generation process.

\begin{table*}[t!]
{\small 
\centering
\begin{tabular}{ccccccc} &\\ [-.3cm]  
$p$ & $\longrightarrow$ & 10 & $20$ & $30$ & $40$ & $50$ 
\tabularnewline & \\ [-.3cm] \hline &\\ [-.3cm]
$6 p^3$ & $\longrightarrow$ & 6,000 & $ 48,000 $ & $ 162,000 $ & $ 384,000 $ & $ 750,000$ 
\tabularnewline & \\ [-.3cm] \hline &\\ [-.3cm]
$r_{vr}$ & $\longrightarrow$ & 0.717640 & 0.717640 & 0.717641 & 0.717640 & 0.717640   
\tabularnewline &\\ [-.3cm] \hline &\\ [-.3cm] 
$r_{jl}$ & $\longrightarrow$ & 0.824084 & 0.819960 & 0.818755 & 0.818185 & 0.817809   
\tabularnewline &\\ [-.3cm] \hline &\\ [-.3cm]
\end{tabular}
\caption{Quality of meshes of a unit sphere measured by the metrics (\ref{VR}) and (\ref{JL}) as $p$ increases} 
}
\label{table1}
\end{table*}
  
\begin{table*}[t!]
{\small 
\centering
\begin{tabular}{cccccccc} &\\ [-.3cm]  
$a$ & $\longrightarrow$ & 1.0 & 0.8 & 0.6 & 0.4 & 0.2 & 0.1   
\tabularnewline & \\ [-.3cm] \hline &\\ [-.3cm]
$r_{vr}$ for $p=10$ & $\longrightarrow$ & 0.717640 & 0.749023 & 0.573403 & 0.390018 & 0.201354  & 0.112168   
\tabularnewline &\\ [-.3cm] \hline &\\ [-.3cm] 
$r_{vr}$ for $p=50$ & $\longrightarrow$ & 0.717640 & 0.748662 & 0.572133 & 0.387183 & 0.194082  & 0.097374    
\tabularnewline &\\ [-.3cm] \hline &\\ [-.3cm]
$r_{jl}$ for $p=10$ & $\longrightarrow$ & 0.824084 & 0.779466 & 0.721367 & 0.659456 & 0.511977  & 0.300437  
\tabularnewline &\\ [-.3cm] \hline &\\ [-.3cm] 
$r_{jl}$ for $p=50$ & $\longrightarrow$ & 0.817809 & 0.773979 & 0.719027 & 0.659455 & 0.511977  & 0.290470    
\tabularnewline &\\ [-.3cm] \hline &\\ [-.3cm]
\end{tabular}
\caption{Measures (\ref{VR}) and (\ref{JL}) of ellipsoid meshes for two values of $p$ and decreasing aspect ratios $a$} 
}
\label{table2}
\end{table*}
		
\begin{table*}[t!]
{\small 
\centering
\begin{tabular}{ccccccc} &\\ [-.3cm]  
$b$ & $\longrightarrow$ & 0.0 & 0.1 & 0.2 & 0.3 & 0.4    
\tabularnewline & \\ [-.3cm] \hline &\\ [-.3cm]
$r_{vr}$ for $p=10$ & $\longrightarrow$ & 0.717640 & 0.664304 & 0.441041 & 0.287242 & 0.181349    
\tabularnewline &\\ [-.3cm] \hline &\\ [-.3cm] 
$r_{vr}$ for $p=50$ & $\longrightarrow$ & 0.717640 & 0.663084 & 0.439356 & 0.285817 & 0.180264     
\tabularnewline &\\ [-.3cm] \hline &\\ [-.3cm]
$r_{jl}$ for $p=10$ & $\longrightarrow$ & 0.824084 & 0.586850 & 0.298063 & 0.185235 & 0.128325   
\tabularnewline &\\ [-.3cm] \hline &\\ [-.3cm] 
$r_{jl}$ for $p=50$ & $\longrightarrow$ & 0.817809 & 0.113584 & 0.052925 & 0.032887 & 0.022714     
\tabularnewline &\\ [-.3cm] \hline &\\ [-.3cm]
\end{tabular}
\caption{Measures (\ref{VR}) and (\ref{JL}) of meshes of $\Omega$ given by (\ref{concav}) for $p=10$, $p=50$ and increasing $b$} 
}
\label{table3}
\end{table*}	

In Table 3 we present the same type of results as in Table 2. However, in contrast to the latter case, $r_{jl}$ now indicates a clear degeneracy of element shapes, as the domain becomes more distorted, i.e. as $b$ increases. This effect is amplified by the discrepancy between values of this parameter as the mesh is refined. 
Nevertheless a rather stable behavior of parameter $r_{vr}$ can be observed. 
  	
\section{Final comments}

\begin{enumerate}
\item 
Some problems may arise when using the mesh generation procedure described in this article, in case the function $f$ defining the boundary of the domain  
has large local Lipschitz constants with respect to the spherical coordinates $\theta$ and $\phi$. A similar situation may happen in the two-dimensional case for 
the triangulation procedure studied in \cite{CMA}. Actually in that paper indications are given on how to remedy eventual "inside-out turning" of elements, which may occur in such cases. However we will not further elaborate on those issues here, since anyway it is not advisable to mesh too distorted domains using our method.
\item
The unknown numbering issue for a discretization method to be used in connection with the tetrahedrization proposed in this work has been examined as well. As one 
can easily guess, it is possible to conceive rather simple algorithms for optimal unknown or node numbering using the analogies with the unit cube. However for the sake of brevity we skip details. 
\item
As one can easily infer from the description and the examples given in the previous sections, a natural by-product of our mesh generation procedure is a family 
of quasi-uniform triangulations of the surface of spheroidal domains indexed by the mesh parameter $p$, in case it is not too distorted. This kind of mesh is 
very useful in shell modeling and in CAD, among other applications. 
\item 
Local mesh refinement is possible with the procedure proposed in this paper. It suffices to start from a locally refined mesh of the unit cube, and then map the resulting vertex coordinates into the true curved domain in the way prescribed in Sections 2 and 3. However our method is not well adapted to such refinements because after all it only generates structured meshes in a certain sense. This means that local refinement necessarily impacts zones far away from the one where it is necessary, likewise finite difference grids. Moreover the number of mesh elements and nodes must remain constant for a given value of the mesh integer  parameter $p$, and therefore local refinement necessarily implies mesh coarsening away from the refined zone. As a corollary, our mesh generation method is unsuitable to adaptivity techniques. Nevertheless it is certainly very useful whenever one is dealing with problems having a smooth solution in a curved domain not so irregular. In this case  the user can take the best advantage of these features, by avoiding low quality meshes that might result from general meshing algorithms. 
\item
The procedure studied in this paper was first proposed by the author in two papers quoted in \cite{cmame2017}, published in the 80's. 
One of them written in Portuguese appeared in Revista Brasileira de Computa\c{c}\~ao; the other one was its abridged translation into English published in a  conference proceedings. However, to the best of author's knowledge, this procedure was implemented for the first time in \cite{cmame2017} and had not been the object of any assessment prior to the present work.                            
\end{enumerate}
 
\noindent \underline{Acknowledgment:}
The author is grateful to CNPq for the financial support through grant 307996/2008-5.

\end{document}